\documentstyle[12pt]{article}
\newcommand\be{\begin{equation}}
\newcommand\ee{\end{equation}}
\newcommand\bea{\begin{eqnarray}}
\newcommand\eea{\end{eqnarray}}
\newcommand\ket[1]{|#1\rangle}
\newcommand\bra[1]{\langle #1|}

\newcommand{\fatalpha}{{\bf \alpha \kern -0.44em \alpha}}
\newcommand{\fatsigma}{{\bf \sigma \kern -0.54em \sigma}}
\newcommand{\tpchi}{{\bf \chi \kern -0.35em \chi}}
\newcommand{\llambda}{{\bf \lambda \kern -0.45em \lambda}}

\renewcommand{\theequation}{\arabic{equation}}
\renewcommand{\theequation}{\thesection-\arabic{equation}}
\bibliography{plain}
\title{\bf Optimal Unambiguous
Discrimination of Quantum States} \vspace{20mm}
\author{ M. A.  Jafarizadeh$^{a,b,c}$
 \thanks{E-mail:Jafarizadeh@tabrizu.ac.ir},
 M. Rezaei$^{b}$ \thanks{E-mail:Karamaty@tabrizu.ac.ir}, N. Karimi$^{a}$ \thanks{E-mail:karimi@tabrizu.ac.ir}, A. R. Amiri$^{a}$ \thanks{E-mail:amiri@tabrizu.ac.ir}\\
\\
$^a${\small The Department of Theoretical Physics and
Astrophysics,
 University of Tabriz, Tabriz, 51664, Iran.} \\
$^b${\small The Institute for Studies in Theoretical Physics and
Mathematics,
Tehran 19395-1795, Iran.} \\
$^c${\small The Research Institute for Fundamental Sciences,
Tabriz 51664, Iran.}} \pagebreak

\vspace{20mm}
\begin{document}
\maketitle \vspace{15mm}
\newpage
\begin{abstract}
Unambiguously distinguishing between nonorthogonal but linearly
independent quantum states is a challenging problem in quantum
information processing. In this work,  an exact analytic solution to
an optimum measurement problem involving an arbitrary number of pure
linearly independent quantum states is presented. To this end,  the
relevant semi-definite programming task  is reduced to a linear
programming one with a feasible region of polygon type which can be
solved via simplex method. The strength of the method is illustrated
through some explicit examples. Also using the close connection
between the Lewenstein-Sanpera decomposition(LSD) and semi-definite
programming approach,  the optimal positive operator valued measure
for some of the well-known examples is  obtain via
Lewenstein-Sanpera decomposition method.
{\bf PACs Index: 03.67.Hk, 03.65.Ta, 42.50.-p

 }
\end{abstract}
\pagebreak
\newpage
\section{Introduction}
In quantum information and quantum computing, the carrier of
information is some quantum system and information is encoded in
its state. A quantum state describes what we know about a quantum
system. Given a single copy of a quantum system which can be
prepared in several known quantum states, our aim is to determine
in which state the system is. This can be well-understood in a
communication context where only a single copy of the system is
given and only a single shot-measurement is performed. This is in
contrast with usual experiments in physics where many copies of a
system are measured to get the probability distribution of the
system. In quantum state discrimination, no statistics is built
since only a single-shot measurement is performed on a single copy
of the system. Actually there are fundamental limitations to the
precision with which the state of the system can be determined
with a single measurement. A fundamental aspect of quantum
information theory is that non-orthogonal quantum states cannot be
perfectly distinguished. Therefore, a central problem in quantum
mechanics is to design measurements optimized to distinguish
between a collection of non-orthogonal quantum states. The topic
of quantum state discrimination was firmly established in 1970s by
pioneering work of Helstrom \cite{helstrom}, who considered a
minimum error discrimination of two known quantum states. In this
case, the state identification is probabilistic. Another possible
discrimination strategy is the so-called unambiguous state
discrimination (USD) where the states are successfully identified
with non-unit probability, but without error. USD was originally
formulated and analyzed by Ivanovic, Dieks and Peres
\cite{ivanovic,dieks,peres} in 1987. The solution for unambiguous
discrimination of two known pure states appearing with arbitrary
prior probabilities was obtained by Jaeger and
Shimony\cite{shimony}. Although the two-state problem is well
developed, the problem of unambiguous discrimination between
multiple quantum states has received considerably less attention.
The problem of discrimination among three nonorthogonal states was
first considered by Peres and Terno
   \cite{peres}. They developed a geometric approach and applied
   it numerically on several examples. A different method was
    considered by Duan and Guo \cite{duan} and Y. Sun and et al \cite{sun}.
Chefles \cite{chef} showed that a necessary and sufficient
condition for the existence of unambiguous measurements for
distinguishing between $N$ quantum states is that the states are
linearly independent. He also proposed a simple suboptimal
measurement for unambiguous discrimination for which the
probability of an inconclusive result is the same regardless of
the state of the system. Equivalently, the measurement yields an
equal probability of correct detection of each one of the ensemble
states.

Over the past years, semidefinite programming (SDP) has been
recognized as a valuable numerical tool for control system analysis
and design. In SDP, one minimizes a linear function subject to the
constraint that an affine combination of symmetric matrices is
positive semidefinite. SDP has been studied (under various names) as
far back as the 1940s. Subsequent research in semidefinite
programming during the 1990s was driven by applications in
combinatorial optimization\cite{Goemans}, communications and signal
processing \cite{luo,davidson,ma}, and other areas of
engineering\cite{se}. Although semidefinite programming is designed
to be applied in numerical methods, it can be used for analytic
computations, too. In the context of quantum computation and quantum
information, Barnum, Saks and Szegedy have reformulated quantum
query complexity and in terms of a semidefinite program \cite{saks},
 while  M. A. Jafarizadeh, M. Mirzaee and M. Rezaee have revealed the
close connection between  Lewenstein-Sanpera decomposition,
robustness of entanglement, finite quantum tomography  and
semi-definite programming algorithm.
\cite{jafarizadeh1,jafarizadeh7, jafarizadeh8,jafarizadeh9}.

The problem of finding the optimal measurement to distinguish
between a set of quantum states was first formulated as a
semidefinite program in 1972 by Holevo, who gave optimality
conditions equivalent to the complementary slackness conditions
\cite{helstrom}. Recently, Eldar, Megretski and Verghese showed that
the optimal measurements can be found efficiently by solving the
dual followed by the use of linear programming \cite{eldarsdp}. Also
in \cite{shor}, SDP has been used to show that the standard
algorithm implements the optimal set of measurements. All of the
above mentioned applications indicate that the method of SDP is very
useful. The reason why the area has shown relatively slow progress
until recently within the rapidly evolving field of quantum
information is that it poses quite formidable mathematical
challenges. Except for a handful of very special cases, no general
exact solution has been available involving more than two arbitrary
states and mostly numerical algorithms are proposed for finding
optimal measurements for quantum-state discrimination, where the
theory of the semi-definite programming provides a simple check of
the optimality of the numerically obtained results.
\par
In this study, we obtain the feasible region in terms of the inner
product of the states which enables us to solve the problem
analytically without using dual states. Exact analytical solution
for optimal unambiguous discrimination of $N$ linearly independent
pure states is calculated and a geometrical interpretation for
minimum inconclusive result for unambiguous discrimination of two
pure states is presented. For more than three states, the analytical
calculations is very complicated to write down and therefore we will
consider the spacial cases such as geometrical uniform states and
Welch bound equality (WBE) sequences. To solve the problem in
general form, following prescription of Refs
\cite{jafarizadeh2,jafarizadeh3,jafarizadeh4,jafarizadeh5,jafarizadeh6},
we have reduced it to LP one, where the computation can be done at a
very fast pace and with high precision. Moreover, we obtain the
feasible region in terms of the inner product of the dual states and
show that LSD is equivalent to optimal unambiguous state
discrimination, and thus one can use LSD to solve the problem of
optimal unambiguous state discrimination. This method is illustrated
for two and three linearly independent states explicitly.
\par
The organization of the paper is as follows: First the definition of
the unambiguous quantum state discrimination are given. After that
in section 3 unambiguous discrimination of quantum states by
introducing  feasible region and using linear programming are
discussed. Then Lewnstein-sanpera decomposition is studied as an
optimal unambiguous discrimination of quantum states. Finally,
discrimination of non-orthogonal quantum states using approximated
linear programming are discussed. The paper is ended with a brief
conclusion and an appendix.
\section{Unambiguous quantum state discrimination}
In quantum theory, measurements are represented by positive
operator valued measures (POVMs). A measurement is described by a
collection $\{M_{k}\}$ of measurement operators. These operators
are acting on the state space of the system being measured. The
index $k$ refers to the measurement outcomes that may occur in the
experiment. In quantum information theory the measurement
operators $\{M_{k}\}$ are often called Kraus operators
\cite{krus}. If we define the operator \be \Pi_{k}=
M_{k}^{\dagger}M_{k}, \ee the probability of obtaining the outcome
$k$ for a given state $\rho_{i}$ is given by $p(k|i)
=Tr(\Pi_{k}\rho_{i})$. Thus, the set of operators $\Pi_{k}$ is
sufficient to determine the measurement statistics.

 {\bf Definition (POVM)}. A set of operators $\{\Pi_{k}\}$ is named a
positive operator valued measure (POVM) if and only if the
following two conditions are met:
\begin{enumerate}
    \item Each operator $\Pi_{k}$ is positive $\;\;\;\Leftrightarrow \;\;\
\langle \psi\mid \Pi_{k} \mid \psi\rangle \geq  0,\;\;\ \forall
\;\ \mid \psi\rangle.$
    \item The completeness relation  is satisfied, i.e.,
\be\label{po2}\sum_{k} \Pi_{k} = 1.\ee
\end{enumerate}
The elements of $\{\Pi_{k}\}$ are called effects or POVM elements.
On its own, a given POVM $\{\Pi_{k} \}$ is enough to give complete
knowledge about the probabilities of all possible outcomes;
measurement statistics is the only item of interest. Consider a
set of known states ${\rho_{i}},i=1,...,N,$ with their prior
probabilities ${\eta_{i}}$. We are looking for a measurement that
either identifies a state uniquely (conclusive result) or fails to
identify it (inconclusive result). The goal is to minimize the
probability of inconclusive result. The measurements involved are
typically generalized measurements. A measurement described by a
POVM $\{\Pi_{k}\}_{k=1}^N$ is called unambiguous state
discrimination measurement(USDM) on the set of states
$\{\rho_{i}\}_{i=1}^N$ if and only if the following conditions are
satisfied:\\

\begin{enumerate}
    \item The POVM contains the elements $\{\Pi_{k}\}_{k=0}^N$ where $N$
is the number of different signals in the set of states .The
element $\Pi_{0}$ is related to an inconclusive result, while the
other elements correspond to an identification of one of the
states $\rho_{i}$, $i=1,...,N$.
    \item No states are wrongly identified, that is,
$Tr(\rho_{i}\Pi_{k})=0\quad\quad \forall i\neq k\quad
i,k=1,...,N.$
\end{enumerate}

 Each USD measurement gives rise to a failure probability,
that is, the rate of inconclusive result. This can be calculated
as
\begin{equation}
Q=\sum_{i}\eta_{i}Tr(Tr(\rho_{i}\Pi_{0})).
 \end{equation}
 And the success probability can be calculated as
 \begin{equation}
P=1-Q=\sum_{i}\eta_{i}Tr(Tr(\rho_{i}\Pi_{i})).
 \end{equation}
 A measurement described by a POVM $\{\Pi_{k}^{opt}\}$
 is called an optimal unambiguous state discrimination
measurement (OptUSDM) on a set of states $\{\rho_{i}\}$ with the
corresponding prior probabilities $\{\eta_{i}\}$ if and only if
the following conditions are satisfied:
\begin{enumerate}
    \item The POVM $\{\Pi_{k}^{opt}\}$ is a USD measurement on $\{\rho_{i}\}$
    \item The probability of inconclusive result is minimal, that is,
$Q(\{\Pi_{k}^{\mathrm{opt}}\}) = \min Q(\{\Pi_{k}\})$ where the
minimum is taken over all USDM.
\end{enumerate}
 Unambiguous state discrimination is an
error-free discrimination. This implies a strong constraint on the
measurement. Suppose that a quantum system is prepared in a pure
quantum state drawn from a collection of given states
$\{|\psi_{i}\rangle,1\leq i\leq N\}$ in $d$-dimensional complex
Hilbert space $\mathcal{H}$ with $d\geq N$. These states span a
subspace $\mathcal{U}$ of $\mathcal{H}$. In order to detect the
state of the system, a measurement is constructed comprising $N+1$
measurement operators $\{\Pi_{i},0\leq i\leq N\}$. Given that the
state of the system is $|\psi_{i}\rangle$, the probability of
obtaining outcome $k$ is
$\langle\psi_{i}|\Pi_{k}|\psi_{i}\rangle$. Therefore, in order to
ensure that each state is either correctly detected or an
inconclusive result is obtained, we must have
\begin{equation}\label{unamb}
\langle\psi_{i}|\Pi_{k}|\psi_{i}\rangle=p_{i}\delta_{ik},\quad
1\leq i,k \leq N,
\end{equation}
for some $0\leq p_{i}\leq 1$. Since $ \Pi_{0} =
I_{d}-\sum^{N}_{i=1}\Pi_{i}$, we have
$\langle\psi_{i}|\Pi_{0}|\psi_{i}\rangle=1-p_{i}$. So a system
with given state $|\psi_{i}\rangle$, the state of the system is
correctly detected with probability $p_{i}$ and an inconclusive
result is obtained with probability $1-p_{i}$. It was shown in
\cite{chef} that (\ref{unamb}) is satisfied if and only if the
vectors $|\psi_{i}\rangle$ are linearly independent, or
equivalently, $\dim \mathcal{U}=\dim\mathcal{H}$. Therefore, we
will take this assumption throughout the paper. In this case, we
may choose \cite{eld1}
\begin{equation}
\Pi_{i}=p_{i}\ket{\tilde{\psi_{i}}}\bra{\tilde{\psi_{i}}},\quad
1\leq i\leq N,
\end{equation}
where, the vectors $|\tilde{\psi}_{i}\rangle \in \mathcal{U}$ are
the reciprocal states associated with the states
$|\psi_{i}\rangle$, i.e., there are unique vectors in
$\mathcal{U}$ such that
\begin{equation}
\langle\tilde{\psi_{i}}|\psi_{k}\rangle=\delta_{ik},\quad 1\leq
i,k\leq N.
\end{equation}
With $\Phi$ and $\tilde{\Phi}$ we denote the matrices such that
their columns are $|\psi_{i}\rangle$ and
$|\tilde{\psi}_{i}\rangle$, respectively. Then, one can show that
$\tilde{\Phi}$ is
\begin{equation}
\tilde{\Phi}=\Phi(\Phi\Phi^{\ast})^{-1}.
\end{equation}
Since the vectors $|\psi_{i}\rangle$, $i=1,...,N$ are linearly
independent, $ \Phi\Phi^{\ast} $ is always invertible.
Alternatively,
\begin{equation}
\tilde{\Phi}=(\Phi\Phi^{\ast})^{\ddag}\Phi,
\end{equation}
so that
\begin{equation}
\ket{\tilde{\psi_{i}}}=(\Phi\Phi^{\ast})^{\ddag}\ket{{\psi_{i}}}
\end{equation}
 where, $(.)^{\ddag}$ denotes the Moore-Penrose pseudo-inverse
\cite{golub}. The inverse is taken on the subspace spanned by the
columns of the matrix. If the state $|\psi_{i}\rangle$ is prepared
with prior probability $\eta_{i}$, then the total probability of
correctly detecting the state is
\begin{equation}
P=\sum^{N}_{i=1}\eta_{i}\langle\psi_{i}|\Pi_{i}|\psi_{i}\rangle=\sum^{N}_{i=1}\eta_{i}p_{i}
\end{equation}
and the probability of the inconclusive result is given by
\begin{equation}
Q=1-P=\sum^{N}_{i=1}\eta_{i}\langle\psi_{i}|\Pi_{0
}|\psi_{i}\rangle=1-\sum^{N}_{i=1}\eta_{i}p_{i}.
 \end{equation}
 In general, an optimal measurement for a given strategy depends on the quantum states and
the prior probabilities of their appearance. In the unambiguous
discrimination for a given strategy and a given ensemble of states,
the goal is to find a measurement which minimizes the inconclusive
result. In fact, it is known that USD (of both pure and mixed
states) is a convex optimization problem . Mathematically, this
means that the quantity which is to be optimized as well as the
constraints on the unknowns, are convex functions. Practically, this
implies that the optimal solution can be computed in an extremely
efficient way. This is therefore a very useful tool. Nevertheless
our aim is to understand the structure of USD in order to relate it
with neat and relevant quantities and to find feasible region for
numerical and analytical solutions. So, by using SDP we determine
feasible region via reciprocal states and reduce the theory of the
SDP to a linear programming one with a feasible region of polygon
type which can be solved via simplex method exactly or
approximately.
\section{Unambiguous discrimination of quantum states using linear programming }
The method presented in this section, seems to be a powerful
method which enables us to analytically discriminate $N$ linearly
independent pure quantum states.
 We naturally want to the probabilities $p_{i}$ be as
large as possible in order to increase the detection probabilities,
but their values are bounded by the demand of positivity of
$\Pi_{0}$.
 Let
$$\ket{\psi}:=\sum_{i=1}^{N}a_{i}\ket{\psi_{i}}+\ket{\psi^{\perp}}$$
with $\langle\psi_{i}\mid\psi^{\perp}\rangle=0$, and normalization
condition is defined by \be \label{xx}
\langle\psi\mid\psi\rangle=\sum_{i,j=1}^{N}a_{i}^{\ast}a_{j}\langle\psi_{i}\mid\psi_{j}\rangle
+\langle\psi^{\perp}\mid\psi^{\perp}\rangle=1 .\ee The relation
(\ref{xx}) can be written as \be\label{gram1}
\sum_{i,j=1}^{N}a_{i}^{\ast}a_{j}G_{ij}\leq 1, \ee where,
$G_{ij}=\langle\psi_{i}\mid\psi_{j}\rangle$ are matrix elements of
the Gram matrix. Now we define the following vector representation
\be X:= \left(\begin{array}{c}
                                                        a_{1} \\
                                                           \vdots \\
                                                           a_{n} \\
                                                         \end{array}\right),\ee
then, the Eq. (\ref{gram1}) can be rewritten as the following
constraint \be X^{\dagger}G X \leq 1.\ee
 Positivity of $Tr(\Pi_{0})$ gives
 $$
 \sum_{i}a_{i}^{2}p_{i}\leq 1.
 $$
 This last condition is a decisive one that actually determines
 the domain of acceptable values of $p_{i}$.
 This result leads us to
the optimization problem defined as
\be
\begin{array}{c}
  \hspace{1cm}\mbox{maximize}\;\;\sum_{i}\eta_{i}p_{i}\leq 1 \\
    \mbox{s.t}\;\;\left\{\begin{array}{c}
        \sum_{i}a^{2}_{i}p_{i} \leq 1 \\
        X^{\dagger}G X \leq 1 .\\
      \end{array}\right.
\end{array}
\ee
   If we write $p_{i}=\lambda^{'} \xi_{i}$ with $0\leq \xi_{i}\leq 1$, then we will have
   \begin{equation}
\sum_{i}a^{2}_{i}p_{i}=\lambda^{'} \sum_{i}a^{2}_{i}\xi_{i}\leq 1
\Rightarrow \lambda^{'}\leq \frac{1}{\sum_{i}a^{2}_{i}\xi_{i}}.
    \end{equation}
   Then, we must compute the maximum value of
    $\sum_{i}a^{2}_{i}\xi_{i}$ such that $\lambda^{'}$ possesses its
    lowest possible value. That is,
\be
\begin{array}{c}
     \hspace{-1cm} \mbox{maximize}\;\; \sum_{i}a^{2}_{i}\xi_{i}\\
     \hspace{-1cm}\mbox{s.t}\;\;\;X^{\dagger}G X \leq 1. \\
   \end{array}\ee
By defining a diagonal matrix as follows \be
D:=\left(\begin{array}{cccc}
 \xi_{1} & 0 & \ldots & 0 \\
  0 & \xi_{2} & \ldots & 0 \\
  \vdots & \vdots & \ddots & \vdots \\
  0 & 0 & \cdots & \xi_{d} \\
\end{array}\right),
\ee
 we can write $\sum_{i}a^{2}_{i}\xi_{i}=X^{\dagger}D X$.
 This leads us to the following optimization problem
\be\label{xx'}
\begin{array}{c}
       \hspace{-1cm}\mbox{maximize}\;\; X^{\dagger}DX\\
     \hspace{-1cm}\mbox{s.t}\;\;X^{\dagger}GX  \leq 1. \\
   \end{array}\ee
Now, let $Y=\sqrt{D}X$. Then, (\ref{xx'}) can be rewritten as
 \be\label{gram4}
\begin{array}{c}
       \hspace{-2.7cm}\mbox{maximize} \;\;Y^{\dagger} Y\\
     \hspace{-1cm}\mbox{s.t}\;\;Y^{\dagger}D^{-\frac{1}{2}}G D^{-\frac{1}{2}} Y \leq 1. \\
   \end{array}\ee
   Suppose $D^{-\frac{1}{2}}G D^{-\frac{1}{2}}=\widehat{G}$. Then,
   we have
   \be
\begin{array}{c}
       \hspace{-1cm}\mbox{maximize}\;\;Y^{\dagger}Y\\
    \hspace{-1cm} \mbox{s.t}\;\;Y^{\dagger}\widehat{G}Y  \leq 1. \\
   \end{array}\ee
 The determinant of $\tilde{G}-\lambda I=0$ determines the feasible
region provided that, $\lambda$ coincides with $\lambda^{'}$. Now,
in order to show the ability of our method, we calculate the optimal
failure probabilities corresponding to unambiguous discrimination of
two and three linearly independent states with arbitrary prior
probabilities.
  In the simple case of two pure states $|\psi_{1}\rangle$ and $|\psi_{2}\rangle$
 with arbitrary prior probabilities $\eta_{1}$ and $\eta_{2}$, $\widehat{G}$ is given by
  \begin{equation}
  \widehat{G}=\left(
\begin{array}{cc}
  \frac{1}{\sqrt{\xi_{1}}} & 0 \\
   0& \frac{1}{\sqrt{\xi_{2}}} \\
\end{array}
\right)\left(
\begin{array}{cc}
 1& a_{12} \\
  a^{\ast}_{12} & 1 \\
\end{array}\right)\left(
\begin{array}{cc}
  \frac{1}{\sqrt{\xi_{1}}} & 0 \\
   0& \frac{1}{\sqrt{\xi_{2}}} \\
\end{array}
\right)=\left(
\begin{array}{cc}
  \frac{1 }{\xi_{1}}& \sqrt{\frac{1}{\xi_{1}(\xi_{2})}}a_{12} \\
  \sqrt{\frac{1}{\xi_{1}(\xi_{2})}} a^{\ast}_{12}& \frac{1}{\xi_{2}} \\
\end{array}
\right).
  \end{equation}
The characteristic equation is given by
\begin{equation}
\lambda^{2}-(\frac{1}{\xi_{1}}+\frac{1}{\xi_{2}})\lambda+\frac{1}{\xi_{1}\xi_{2}}(1-|a^{2}_{12}|)=0.
\end{equation}
Thus,
\begin{equation}\label{x}
\xi_{1}\xi_{2}\lambda^{2}-\lambda(\xi_{1}+\xi_{2})+(1-|a^{2}_{12}|)=0.
\end{equation}
In the feasible region we have $p_{1}=\lambda\xi_{1}$ and
$p_{2}=\lambda\xi_{2}$, then the Eq. (\ref{x}) is equivalent to
\begin{equation}\label{fise1}
 p_{1}p_{2}-(p_{1}+p_{2})+(1-|a^{2}_{12}|)=0.
\end{equation}
This equation  determines the feasible region (see Figure. $1$).
To calculate the minimum probability of inconclusive result we put
equal the gradient of line $\eta_{1}p_{1}+\eta_{2}p_{2}$ to the
gradient of equation (3-27), then we will have
\begin{equation}
p_{2}-1=\Lambda\eta_{1}\quad and\quad p_{1}-1=\Lambda\eta_{2}
\end{equation}
By substituting the equation (3-28) into (3-27), we obtain
\begin{equation}
(1+\Lambda\eta_{1})(1+\Lambda\eta_{2})-(2+\Lambda)+(1-a^{2}_{12})=0,
\end{equation}
which implies that
$\Lambda=\pm\frac{|a_{12}|}{\sqrt{\eta_{1}\eta_{2}}}$. Substituting
$\Lambda$ into equation (3-28), gives the following solutions:
\begin{equation}
p_{1}=1\pm \sqrt{\frac{\eta_{2}}{\eta_{1}}}|a_{12}|\quad and \quad
p_{2}=1\pm \sqrt{\frac{\eta_{1}}{\eta_{2}}}|a_{12}|.
\end{equation}
Since $p_{i}\leq 1$, thus we conclude that
$$
p_{1}=1-\sqrt{\frac{\eta_{2}}{\eta_{1}}}|a_{12}|\quad and \quad
p_{2}=1- \sqrt{\frac{\eta_{1}}{\eta_{2}}}|a_{12}|.
$$
From the positivity of $p_{1}$ and $p_{2}$, we have:
$$
|a_{12}|\leq \sqrt{\frac{\eta_{2}}{\eta_{1}}}\leq
\frac{1}{|a_{12}|}.
$$
In this case, the minimum  probability of inconclusive result is:
\begin{equation}
Q=1-(\eta_{1}p_{1}+\eta_{2}p_{2})=
2\sqrt{\eta_{1}\eta_{2}}|a_{12}|.
 \end{equation}
If $\sqrt{\frac{\eta_{2}}{\eta_{1}}}<|a_{12}|$, then $p_{2}=0$ and
if $\frac{1}{|a_{12}|}<\sqrt{\frac{\eta_{2}}{\eta_{1}}}$, then
$p_{1}=0$. Then, by using (\ref{fise1}), we obtain \be\left\{
\begin{array}{c}\label{dotae}
    \hspace{-1cm}  Q=\eta_{2}+\eta_{1}|a_{12}|^{2} \quad if \quad
\sqrt{\frac{\eta_{2}}{\eta_{1}}}< |a_{12}|\\
 Q=2\sqrt{\eta_{1}\eta_{2}}|a_{12}|\quad if \quad |a_{12}|\leq
\sqrt{\frac{\eta_{2}}{\eta_{1}}}\leq \frac{1}{|a_{12}|} \\
\hspace{-1cm} Q=\eta_{1}+\eta_{2}|a_{12}|^{2} \quad if \quad
\frac{1}{|a_{12}|}<\sqrt{\frac{\eta_{2}}{\eta_{1}}}
   \end{array}\right.\ee
    Here, we  discuss geometrical interpretation of optimal unambiguous discrimination of two pure
   states on Bloch sphere. One can show that the minimum inconclusive result for unambiguous discrimination
    of two pure states is
   equivalent to distance between sphere center and the line
   connecting $\rho_{1}$ to $\rho_{2}$ (see Figure $2$).
   Density matrix for a pure qubit state is defined in the Bloch form as
follows
\be\begin{array}{c}\rho_{1}=\frac{1}{2}(I_{2}+{n_{1}}. \sigma)\\
\rho_{2}=\frac{1}{2}(I_{2}+{n_{2}}. \sigma).\end{array} \ee For
unambiguous discrimination we will have \be\label{qubit1x}
\begin{array}{cc} Tr(\Pi_{1}\rho_{2})\rightarrow 0,&  Tr(\Pi_{2}\rho_{1})\rightarrow
0\\ Tr(\Pi_{1}\rho_{1})\rightarrow 1,&
Tr(\Pi_{2}\rho_{2})\rightarrow 1,
\end{array}
\ee where $\Pi_{1}$ and $\Pi_{2}$ are  the POVM elements in the pure
Bloch form
\be\begin{array}{c}\Pi_{1}=\frac{1}{2}(I_{2}+{n_{1}^{\prime}}.
\sigma) \\
\Pi_{2}=\frac{1}{2}(I_{2}+{n_{2}^{\prime}}. \sigma).\end{array} \ee
This is clear from (\ref{qubit1x}) that $Tr(\Pi_{1}\rho_{2})$ is
minimal if ${n_{1}^{\prime}}. n_{2}<0$. Then the optimal case is
attained for ${n_{1}^{\prime}}=-{n_{2}}$ and $Tr(\Pi_{2}\rho_{1})$
is minimal if ${n_{2}^{\prime}}=-{n_{1}}$ then the POVM elements are
given by
\be\begin{array}{c}\Pi_{1}=\frac{1}{2}(I_{2}-{n_{2}}.
\sigma)=\mid \tilde{\psi}_{1}\rangle\langle\tilde{\psi}_{1}\mid, \\
\Pi_{2}=\frac{1}{2}(I_{2}-{n_{1}}. \sigma)=\mid
\tilde{\psi}_{2}\rangle\langle\tilde{\psi}_{2}\mid.\end{array} \ee
For $\eta_1=\eta_2=1/2$, using (\ref{dotae})  optimum $p$'s
coefficients corresponds to $\phi_{1}=\phi_{2}$ and  $p$'s are
given by \be p_{1}=p_{2}=1-|a_{12}|=1-\cos\theta, \ee where
$2\theta=n_{1}.n_{2}$ and then minimum inconclusive result is
equal $Q= \cos\theta$.

Regarding to figure 2 we see that $R=\cos\theta=Q$. That is the
minimum inconclusive result for unambiguous discrimination
    of two pure states is
   equivalent to distance between sphere center and the line which
   connecting $\rho_{1}$ to $\rho_{2}$.

Now, we give analytical solution for three linearly independent
normalized state vectors $|\psi_{1}\rangle$ , $|\psi_{2}\rangle$
and $\psi_{3}$ in the three-dimensional complex vector space with
arbitrary prior probabilities $\eta_{1}$, $\eta_{2}$ and
$\eta_{3}$. If we consider
$a_{ij}=\langle\psi_{i}|\psi_{j}\rangle$, then $\widehat{G}$ is
given by
\begin{equation}
\widehat{G}=\left(
\begin{array}{ccc}
  \frac{1}{\sqrt{\xi_{1}}} & 0 &0 \\
 0 & \frac{1}{\sqrt{\xi_{2}}} & 0 \\
  0 & 0 & \frac{1}{\sqrt{\xi_{3}}} \\
\end{array}
\right) \left(
\begin{array}{ccc}
  1& a_{12} & a_{13} \\
  a^{\ast}_{12} & 1 & a_{23} \\
  a^{\ast}_{13}&a^{\ast}_{23}& 1 \\
\end{array}
\right) \left(
\begin{array}{ccc}
  \frac{1}{\sqrt{\xi_{1}}} & 0 &0 \\
 0 & \frac{1}{\sqrt{\xi_{2}}} & 0 \\
  0 & 0 & \frac{1}{\sqrt{\xi_{3}}} \\
\end{array}\right)=
\left(
\begin{array}{ccc}
  \frac{1}{\xi_{1}} & \frac{a_{12}}{\sqrt{\xi_{1}\xi_{2}}} & \frac{a_{13}}{\sqrt{\xi_{1}\xi_{3}}} \\
\frac{a^{\ast}_{12}}{\sqrt{\xi_{1}\xi_{2}}} & \frac{1}{\xi_{2}}  & \frac{a_{23}}{\sqrt{\xi_{2}\xi_{3}}} \\
  \frac{a^{\ast}_{13}}{\sqrt{\xi_{1}\xi_{3}}}  & \frac{a^{\ast}_{23}}{\sqrt{\xi_{2}\xi_{3}}}  & \frac{1}{\xi_{3}} \\
\end{array}
\right)
\end{equation}
 The characteristic equation is given by
 $$
 \lambda^{3}-(\frac{1}{\xi_{1}}+\frac{1}{\xi_{2}}+\frac{1}{\xi_{3}})
 \lambda^{2}+(\frac{1}{\xi_{1}\xi_{2}}+\frac{1}{\xi_{1}\xi_{3}}+\frac{1}
 {\xi_{2}\xi_{3}}-\frac{a^{2}_{12}}{\xi_{1}\xi_{2}}-\frac{a^{2}_{23}}
 {\xi_{2}\xi_{3}}-\frac{a^{2}_{13}}{\xi_{1}\xi_{3}})\lambda
 $$
 \begin{equation}
 +\frac{1}
 {\xi_{1}\xi_{2}\xi_{3}}(a^{2}_{12}+a^{2}_{23}+a^{2}_{13})-\frac{2a_{12}a_{23}a_{13}}
 {\xi_{1}\xi_{2}\xi_{3}}-\frac{1}{\xi_{1}\xi_{2}\xi_{3}}=0
 \end{equation}
 Since for feasible region we have $p_{i}=\lambda \xi_{i}$, we have
 $$
 p_{1}p_{2}p_{3}-(p_{2}p_{3}+p_{1}p_{3}+p_{1}p_{2})+(1-a^{2}_{23})p_{1}
 $$
\begin{equation}\label{fise3}
+(1-a^{2}_{13})p_{2}+(1-a^{2}_{12})p_{3}+a^{2}_{12}+a^{2}_{23}+a^{2}_{13}-2a_{12}a_{22}a_{13}-1=0
\end{equation}
If the gradient of the plane
$\eta_{1}p_{1}+\eta_{2}p_{2}+\eta_{3}p_{3}$ be equal to the gradient
of Eq. (\ref{fise3}), we will have
 \be\left\{
\begin{array}{c}\label{grade3}
      \;\;p_{2}p_{3}-(p_{2}+p_{3})+1-a^{2}_{23}=\Lambda \eta_{1}\\
\;\;\;\;\;\;p_{1}p_{3}-(p_{1}+p_{3})+1-a^{2}_{13}=\Lambda \eta_{2} \\
   \;\;\;\;\;\;p_{1}p_{2}-(p_{1}+p_{2})+1-a^{2}_{12}=\Lambda
   \eta_{3}.
   \end{array}\right.\ee
   By solving the Eqs. (\ref{fise3}) and (\ref{grade3}),
   one can obtain
\be\left\{
\begin{array}{cc}
      \hspace{-.6cm}\Lambda=0,&\Lambda=\frac{a_{13}a_{23}}{\sqrt{\eta_{1}\eta_{2}}}-\frac{a_{12}
   (a_{13}\sqrt{\eta_{1}}+a_{23}\sqrt{\eta_{2}})}{\sqrt{\eta_{1}\eta_{2}\eta_{3}}}\\
\Lambda=-\frac{a^{2}_{23}}{\eta_{1}},&
\Lambda=\frac{a_{13}a_{23}}{\sqrt{\eta_{1}\eta_{2}}}+\frac{a_{12}
   (a_{13}\sqrt{\eta_{1}}+a_{23}\sqrt{\eta_{2}})}{\sqrt{\eta_{1}\eta_{2}\eta_{3}}}\\
   \Lambda=-\frac{a^{2}_{13}}{\eta_{2}}, &\hspace{.2cm}\Lambda=-\frac{a_{13}a_{23}}{\sqrt{\eta_{1}\eta_{2}}}-\frac{a_{12}
   (a_{13}\sqrt{\eta_{1}}-a_{23}\sqrt{\eta_{2}})}{\sqrt{\eta_{1}\eta_{2}\eta_{3}}}\\
   \Lambda=-\frac{a^{2}_{12}}{\eta_{3}},&\hspace{.2cm}\Lambda=-\frac{a_{13}a_{23}}{\sqrt{\eta_{1}\eta_{2}}}+\frac{a_{12}
   (a_{13}\sqrt{\eta_{1}}-a_{23}\sqrt{\eta_{2}})}{\sqrt{\eta_{1}\eta_{2}\eta_{3}}}.\\
      \end{array}\right.\ee
   but, only $\Lambda=0$ and $\Lambda=-\frac{a_{13}a_{23}}{\sqrt{\eta_{1}\eta_{2}}}+\frac{a_{12}
   (a_{13}\sqrt{\eta_{1}}-a_{23}\sqrt{\eta_{2}})}{\sqrt{\eta_{1}\eta_{2}\eta_{3}}}$
   give the acceptable values for $p_{i}$.
   If the point of contact lies in the first octant and $0\leq
   p_{i}\leq 1 $, it gives the optimal solution. If not, then an
   optimal contact point occurs on one of the
   coordinate planes or even at one of the vertices.

 {\bf Example 1:} We assume that all of the prior
   probabilities are the same and equal to $1/3$. If all of
   the overlaps are the same, i.e., $a_{12}=a_{13}=a_{13}=s$
   where $s$ a is real and positive number, then by using equations (3-35) and (3-36) we will
   obtain
      \be\left\{
\begin{array}{c}
      \;\;p_{1}s^{2}+p_{3}s^{2}+\Lambda/3(1-p_{2})-2s^{2}+2s^{3}=0\\
\;\;\;\;\;\;p_{1}s^{2}+p_{2}s^{2}+\Lambda/3(1-p_{3})-2s^{2}+2s^{3}=0 \\
   \;\;\;\;\;\;p_{2}s^{2}+p_{3}s^{2}+\Lambda/3(1-p_{1})-2s^{2}+2s^{3}=0
   \end{array}\right.\ee
   In this case,  $\Lambda=0$
   gives the optimal values $p_{1}=p_{2}=p_{3}=1-s$ and
   $Q=s$.

We can generalize the above example by considering optimal
distinguishability for WBE. That is,  we  consider equiangular
tight frame WBE sequences (for example special Grassmanian
frames)\cite{Welch}. Let $\{\psi_{i}\}_{i=1}^{N}$ be an
independent frame sequence such that
$$\langle\psi_{i}\mid\psi_{j}\rangle=s\quad {\mathrm{for}} \quad i\neq j$$
\begin{equation}
\langle\psi_{i}\mid\psi_{j}\rangle=1\quad {\mathrm{for}} \quad i=j
\end{equation}
   Thus, for optimal distinguishing between $N$
independent vectors which are prepared with equal prior
probabilities, we can prove (similar to example $1$) that, optimal
$p_i$'s are given by
$$p_{1}=...=p_{N}=1-s.$$
One can also prove that, optimal distinguishability  corresponds
to equal measurement probabilities $p$ that are equal to the
minimum eigenvalue of frame operator. This can be proved by
defining the frame operator as \be
S=\sum_{i=1}^{N}\mid\psi_{i}\rangle\langle\psi_{i}\mid, \ee such
that \be S_{kl}=\sum_{i=1}^{N}\langle
k\mid\psi_{i}\rangle\langle\psi_{i}\mid l\rangle. \ee Then, we
have $S=AA^{\dagger}$ where $A_{ki}=\langle k\mid\psi_{i}\rangle$.
On the other hand we define Gramm matrix as follows \be
G=A^{\dagger}A. \ee One can show that $S$ and $G$ possess equal
eigenvalues, thus we find eigenvalues of Gramm matrix instead of
those of the frame operator. By using (3-45) and (3-48), one can
easily see that for equiangular tight frame, the Gram matrix can
be written as
 \be G=I+s(C-1), \ee where $C$ is a
matrix such that all of its matrix entries are equal to one, thus
we have $C^{2}=NC$. Therefore, eigenvalues of $G$ are
\be
 1+s(N-1),\quad 1-s,
 \ee
 with $s<1$. Then, the minimum eigenvalue of the frame
operator is equal to $1-s$ and thus $p$ is equal to $1-s$.

 {\bf Example 2:} Consider $a_{12}=a_{13}=s_{1}\quad {\mathrm{and}} \quad a_{23}=s_{2}$,
where both $s_{1}$ and $ s_{2}$ are real and positive. Then by
using equations (\ref{fise3}) and (\ref{grade3}) we will have
   \be\left\{
\begin{array}{c}
      \;\;p_{1}s_{2}^{2}+p_{3}s_{1}^{2}+\Lambda/3(1-p_{2})-s_{1}^{2}-s_{2}^{2}+2s_{1}^{2}s_{2}=0\\
\;\;\;\;\;\;p_{1}s_{2}^{2}+p_{2}s_{1}^{2}+\Lambda/3(1-p_{3})-s_{1}^{2}-s_{2}^{2}+2s_{1}^{2}s_{2}=0 \\
   \;\;\;\;\;\;p_{2}s_{1}^{2}+p_{3}s_{1}^{2}+\Lambda/3(1-p_{1})-2s_{1}^{2}+2s_{1}^{2}s_{2}=0.
   \end{array}\right.\ee
By solving the above equations we obtain:
$$
\Lambda=3s^{2}_{1}-6s_{1}s_{2}\Rightarrow
 \left\{
\begin{array}{c}
      \;\;p_{1}=1-2s_{1}\\
\;\;\;\;\;\;p_{2}=p_{3}=s_{2}-s_{1}+1
   \end{array}\right.\Rightarrow Q=2/3(2s_{1}-s_{2}),
$$
\begin{equation}\label{equal}
\Lambda=0\Rightarrow
 \left\{
\begin{array}{c}
      \;\;p_{1}=\frac{s_{2}-s^{2}_{1}}{s_{2}}\\
\;\;\;\;\;\;p_{2}=p_{3}=1-s_{2}
   \end{array}\right.\Rightarrow
   Q=1/3(\frac{s^{2}_{1}}{s_{2}}+2s_{2}).
\end{equation}
One of the $ \Lambda=3s^{2}_{1}-6s_{1}s_{2}$ and $ \Lambda=0$
which gives smaller inconclusive answer, gives the optimal value
provided that lies in the feasible region. If is not, then a
contact point that it is optimal occurs on one of the coordinate
plane or even at one of the vertices. Here we give two numerical
examples such that, in one of them the contact point in equation
(\ref{equal}) is optimal solution, whereas in another one it is
not. First, consider the case in which the ensemble consists of
three linearly independent states with equal prior probabilities
$1/3$, where
\be\begin{array}{ccc}|\psi_{1}\rangle=[1,0,0]^{T}&|\psi_{2}\rangle=\frac{1}{3}[1,2,2]^{T}
& |\psi_{3}\rangle=\frac{1}{3}[1,2,-2]^{T}  .
\end{array}\ee
and $s_{1}=\frac{1}{3}, s_{2}=\frac{1}{9}$. Then, the optimal
solution is given by
\begin{equation}
p_{1}=\frac{1}3{},\quad p_{2}=p_{3}=\frac{7}{9}\quad \mathrm{and}
\quad Q=\frac{10}{27}.
\end{equation}

In figure $3$ we try to show the feasible region of this case
which is a convex region.

 As an another example, we consider the case in which the ensemble
consists of three state vectors with equal probabilities $1/3$,
where
 \be\begin{array}{ccc}|\psi_{1}\rangle=\frac{1}{\sqrt{3}}[1,1,1]^{T}
& |\psi_{2}\rangle=\frac{1}{\sqrt{2}}[1,1,0]^{T} &
|\psi_{3}\rangle=\frac{1}{\sqrt{2}}[0,1,1]^{T}. \end{array}\ee

In this example the contact point occurs on one of the coordinate
plane and optimal solution is given by:
\begin{equation}
p_{1}=0,\quad p_{2}=p_{3}=\frac{1}{6}\quad \mathrm{and} \quad
Q=\frac{1}{9}.
\end{equation}
Figure $4$ shows the feasible region of this example.
\subsection{Equal-probability measurement }
A simple measurement that has been employed for unambiguous state
discrimination is the measurement with $p_{i}=p$,  for all $
i=1,2,...,N$. This measurement results in equal probability of
correctly detecting each of the states and is called
Equal-probability measurement (EPM). Using the feasible region, we
are able to calculate the prior probabilities, so that EPM is
optimal. Using equations (\ref{fise3}) and (\ref{grade3}), the
prior probabilities in the optimal EPM measurement for unambitious
discrimination of three states are given by
$$
\eta_1=\frac{(a^{2}_{12}-1)^{2}}{3-2a^{2}_{12}-a^{2}_{13}-a^{2}_{12}a^{2}_{13}-2a_{23}+2a_{23}a^{2}_{12}+a^{4}_{12}}
$$
$$
\eta_2=\frac{1-a^{2}_{12}-a^{2}_{13}+a^{2}_{12}a^{2}_{13}}
{3-2a^{2}_{12}-a^{2}_{13}-a^{2}_{12}a^{2}_{13}-2a_{23}+2a_{23}a^{2}_{12}+a^{4}_{12}}
$$
\begin{equation}
\eta_3=1-\eta_1-\eta_2.
\end{equation}
In this case the value of $p$ is calculated from equation
(\ref{fise3}). If we consider more than three states, the
functionality of the $\eta_{i}$ in terms of $\{a_{ij}\}$ are too
complicated to be written down, so it is not included here.
However, if we consider geometrically uniform states, the problem
will be easy. Let $S = \{|\psi_i\rangle = U_i|\psi\rangle,U_i \in
{\cal G}\}$ be a set of geometrically uniform (GU) states
generated by a finite group ${\cal G}$ of unitary matrices, where
$|\psi\rangle$ is an arbitrary state. Now, let $\Phi$ be the
matrix with columns $|\psi_{i}\rangle$. Then, the measurement
which minimizes the probability of an inconclusive result could be
reduced to an equal-probability measurement \cite{eld1} and
consists of the measurement operators \be
\Pi_{i}=p_{i}|\tilde{\psi}_{i}\rangle\langle\tilde{\psi}_{i}\mid,
\ee where, $p_{i}$ is the inverse of the maximum eigenvalue of
frame operator \cite{eld1} and
$\mid\tilde{\psi}_{i}\rangle=U_{i}\mid\tilde{\psi}\rangle,\;\;\
U_{i}\in {\cal G}$ with \be
\mid\tilde{\psi}\rangle=(\Phi\Phi^{\ast})^{-1}\mid\psi\rangle,\ee
 In this case, using the feasible region it is easy to show that for
optimal EPM measurement all of prior probabilities are equal. In
general, similar to the example $1$, one can prove that for
optimal EPM measurement all of $p_i$'s are equal to the inverse of
maximum eigenvalue of frame operator.

 {\bf Example3:} We now consider an example of a set of GU
states. Consider the group ${\cal G}$ of $4$ unitary matrices
$U_i$, where \be U_{1}=I_{4},\quad
  U_{2}=\left(\begin{array}{cccc}
  1 & 0 & 0 & 0 \\
  0 & -1 & 0 & 0 \\
  0 & 0 & 1 & 0 \\
  0 & 0 & 0 & -1 \\
\end{array}\right),
\quad U_{3}=\left(\begin{array}{cccc}
  1 & 0 & 0 & 0 \\
  0 & 1 & 0 & 0 \\
  0 & 0 & -1 & 0 \\
  0 & 0 & 0 & -1 \\
\end{array}\right)
,\quad U_{4}=U_{2}U_{3}.
 \ee
 Now, let the set of GU states is given by $S= \{|\psi_{i}\rangle\}=
U_{i}|\psi\rangle, 1 \leq i \leq 4\}$ with $|\psi\rangle =
1/(3\sqrt{2})[2, 2, 1, 3]^{T}$. Then, we obtain \be
\Phi=\frac{1}{3\sqrt{2}}\left(\begin{array}{cccc}
  2 & 2 & 2 & 2 \\
  2 & -2 & 2 & -2 \\
  1 & 1 & -1 & -1 \\
  3 & -3 & -3 & 3 \\
\end{array}\right)
\ee
 It should be noticed that, the reciprocal states
$|\tilde{\psi}_{i}\rangle=U_i|\tilde{\psi}\rangle$ for $i=1,...,4$
with \be
\mid\tilde{\psi}\rangle=(\Phi\Phi^{\ast})^{\dagger}\mid\psi\rangle=\frac{1}{4\sqrt{2}}\left(\begin{array}{c} 3\\ 3\\ 6\\
2\end{array}\right), \ee are also GU states with generating group
${\cal G}$. Therefore, we can provide the elements of POVM as
$\Pi_{i}=p_{i}\mid\tilde{\psi}_{i}\rangle\langle
\tilde{\psi}_{i}\mid$. Using feasible region it easy to show that
the optimal contact point is given by \be
p_{1}=p_{2}=p_{3}=p_{4}=\frac{2}{9}, \ee then the equal
probability measurement operators are given by
\be\Pi_{i}=\frac{2}{9}\mid
\tilde{\psi}_{j}\rangle\langle\tilde{\psi}_{i}\mid,\;\;\ i=1,...,4
\ee where these results are in agreement with those of
Ref.\cite{eld1}.
\section{Lewnstein-Sanpera decomposition (LSD) as an optimal unambiguous discrimination}

 The idea of Refs.
\cite{LS,karnas} is based on the method of subtracting projections
on product vectors from a given state, that is, for a given
density matrix $\rho$ and any set $V=\{|\tilde{\psi}_{i}\rangle\}$
of the states belonging to the range of $\rho$, one can subtract a
density matrix $\rho^\prime=\sum_i p_i \Pi_i$ (not necessarily
normalized) with $p_i\ge 0$ such that $\delta
\rho=\rho-\rho^\prime\ge 0$, in the sense that $Tr(\rho^\prime)\le
1$.

In the following we recall some important definitions and theorems
about LSD.\\
{\bf Definition $1$.} A non-negative parameter $p$ is called
maximal with respect to a (not necessarily normalized) density
matrix $\rho$ and the projection operator $\Pi=\mid
\tilde{\psi}\rangle\langle\tilde{\psi}\mid$ iff $\rho-p \Pi\geq
0$, and for every $\epsilon \geq 0$, the matrix
$\rho-(p+\epsilon)\Pi$ is not positive definite. The maximal $p$
thus determines the maximal contribution of $\Pi$ that can be
subtracted from $\rho$ maintaining the non-negativity of the
difference.

\noindent{\bf Lemma $1$.}  $p$ is  maximal with respect to $\rho$
and $\Pi= |\tilde{\psi}\rangle\langle\tilde{\psi}|$ iff:
\begin{enumerate}
    \item if $|\tilde{\psi}\rangle\not\in {\cal R}(\rho)$ then $p=0$,
    \item if $|\tilde{\psi}\rangle\in {\cal R}(\rho)$ then
\begin{equation}\label{LSD1}
0<p= \frac{1}{\langle
\tilde{\psi}|\frac{1}{\rho}|\tilde{\psi}\rangle}.
\end{equation}
\end{enumerate}
\noindent{\bf Definition $2$.} We say that a pair of non-negative
$(p_1,p_2)$ is  {\it maximal} with respect to $\rho$ and a pair of
projection operators
$\Pi_1=|\tilde{\psi}_1\rangle\langle\tilde{\psi}_1|$,
$\Pi_2=|\tilde{\psi}_2\rangle\langle\tilde{\psi}_2|$ iff
$\rho-p_1\Pi_1-p_2\Pi_2\ge 0$, $p_1$ is maximal with respect to
$\rho-p_2\Pi_2$ and to the projector $\Pi_1$, $p_2$ is maximal
with respect to  $\rho-p_1\Pi_1$ and to the projector $\Pi_2$, and
the sum $p_1+p_2$ is maximal.

\noindent{\bf Lemma $2$.} A pair $(p_1,p_2)$ is maximal with
respect to $\rho$ and a pair of projectors $(\Pi_1,\Pi_2)$ iff:
(a) if $|\tilde{\psi}_1\rangle, \;|\tilde{\psi}_2\rangle$ do not
belong to ${\cal R}(\rho)$ then $p_1=p_2=0$; (b) if
$|\tilde{\psi}_1\rangle$ does not belong to ${\cal R}(\rho)$,
while $|\tilde{\psi}_2\rangle\in {\cal R}(\rho)$ then $p_1=0$,
$p_2=\langle
\tilde{\psi}_2|\frac{1}{\rho}|\tilde{\psi}_2\rangle^{-1}$; (c) if
$|\tilde{\psi}_1\rangle, \;|\tilde{\psi}_2\rangle \in {\cal
R}(\rho)$ and $\langle
\tilde{\psi}_1|\frac{1}{\rho}|\tilde{\psi}_2\rangle=0$ then
$p_i=\langle \tilde{\psi}_i|\frac{1}{\rho}|\tilde{\psi}_i\rangle$,
$i=1,2$; (d) finally,  if $|\tilde{\psi}_1\rangle,
\;|\tilde{\psi}_2\rangle \in {\cal R}(\rho)$ and $\langle
\tilde{\psi}_1|\frac{1}{\rho}|\tilde{\psi}_2\rangle\ne 0$ then
\begin{eqnarray}\label{lsd1}
p_1&=&\frac{1}{D}\left(\langle
\tilde{\psi}_2|\frac{1}{\rho}|\tilde{\psi}_2\rangle-
|\langle \tilde{\psi}_1|\frac{1}{\rho}|\tilde{\psi}_2\rangle|\right), \\
p_2&=&\frac{1}{D}\left(\langle
\tilde{\psi}_1|\frac{1}{\rho}|\tilde{\psi}_1\rangle- |\langle
\tilde{\psi}_1|\frac{1}{\rho}|\tilde{\psi}_2\rangle|\right),
\end{eqnarray}
where $D=\langle
\tilde{\psi}_1|\frac{1}{\rho}|\tilde{\psi}_1\rangle\langle
\tilde{\psi}_2|\frac{1}{\rho}|\tilde{\psi}_2\rangle -|\langle
\tilde{\psi}_1|\frac{1}{\rho}|\tilde{\psi}_2\rangle|^2$.\\
 {\bf Lemma $3$.}
Let a hermitian density matrix $\rho$ has a decomposition of the
form $\rho=\rho^{\prime}+(1-p)\delta \rho$, where $\rho'$ is a
part of density operator $\rho$ which has the structure
$\rho^{\prime}=\sum_{i=1}^{n} p_{i} \Pi_{i},$ with $\Pi_{i}$ being
projection operator onto state $\ket{\tilde{\psi}_{i}}$ and
$\sum_{i=1}^{n} p_{i}=1$. Then the set of $\{ p_{i}\},$ which are
maximal with respect to the density matrix $\rho$ and the set of
the projection operators $\{\Pi_{i}\}$ form a manifold which
generically has dimension $n-1$ and is determined by the following
equation :
\begin{equation}\label{dualfis}
1-\sum_{i}D_{i}p_{i}+\sum_{i<j} D_{ij} p_{i} p_{j}-\sum_{ijk}
D_{ijk}p_{i} p_{j} p_{k}+...+(-1)^{n}
\sum_{i_{1},...,i_{n}}p_{i_{1}}p_{i_{2}}...p_{i_{n}}
D_{i_{1}i_{2}...i_{n}}=0
\end{equation}
 where the set of $\{D_{i_{1}, i_{2}...i_{m}}\}$ are the subdeterminants (minors)
  of  matrix $D$ defined by
\be D=\left(\begin{array}{cccc} \tilde{a}_{11} & \tilde{a}_{12} & ... & \tilde{a}_{1n} \\
\tilde{a}_{21} & \tilde{a}_{22} & ... &\tilde{ a}_{2n} \\ \vdots &
\vdots & \ddots & \vdots \\\tilde{ a}_{n1} &\tilde{ a}_{n2} &
\cdots & \tilde{a}_{nn}
\end{array} \right) ,\ee with $\tilde{a}_{ij}:=\langle\tilde{\psi}_{i}\mid
\frac{1}{\rho}\mid \tilde{\psi}_{j}\rangle$. Equation
(\ref{dualfis}) determines feasible region via reciprocal states,
that is it gives the domain of acceptable values of $p_{i}$. One
way to drive the equation (\ref{dualfis}) is using
semidefinite programming \cite{jafarizadeh1, jafarizadeh8}. \\
 In the rest of this section, we show that optimal unambiguous discrimination for $N$ linearly independent
 states can be reduced to LSD method. Suppose a quantum system is prepared in a state
secretly drawn from a known set $|\psi_{1}\rangle, . . .,
|\psi_{N}\rangle$ where each  $|\psi_{i}\rangle$ is a pure state
in the Hilbert space $\mathcal{H}$. In order to discriminate
$|\psi_{1}\rangle,..., |\psi_{N}\rangle$ unambiguously, one can
construct a most general POVM consisting of $N+1$ elements
$\Pi_{0} ,\Pi_{1}, . . . ,\Pi_{N}$ such that \be\Pi_{i}\geq0\;\
,\;\ i = 0, 1, . . . , N,\;\;\;\mbox{and}\;\;\;\sum_{i=0}^{N}
\Pi_{i} = I,\ee where $I$ denotes the identity matrix in
$\mathcal{H}$. Each element $\Pi_{i}$, $i = 1, . . . , N$ of  POVM
corresponds to an identification of the corresponding state
$|\psi_{i}\rangle$, while $\Pi_{0}$ corresponds to the
inconclusive answer. For the sake of simplicity, we often specify
only $\Pi_{1}, . . . ,\Pi_{N}$ for a given POVM since the left
element $\Pi_{0}$ is uniquely determined by
\be\label{inconcl}\Pi_{0} = I
-\sum_{i=1}^{N}\Pi_{i}=I-\sum_{i}p_{i}|\tilde{\psi}_{i}\rangle\langle\tilde{\psi}_{i}|.\ee
The goal of LSD is maximizing $p_{i}$'s such that $\sum_{i=1}^{N}
p_{i}$ is maximized. With assuming that density matrix in LSD
method is equal to identity we obtain LSD as follows \be \rho-\sum
\Pi_{i}=I-\sum_{i}p_{i}|\tilde{\psi}_{i}\rangle\langle\tilde{\psi}_{i}|.\ee
With comparison this relation and (\ref{inconcl}) it is clear that
we can maximize success probability by using LSD method or
minimize the inconclusive probability is minimized. Then we say
that LSD is the same as OptUSDM and we use LSD in order to obtain
the elements of the optimal POVM.
\subsection{Analytical calculation of optimal POVM for unambiguous
discrimination of quantum states via Lewnstein-Sanpera
decomposition } In this section an analytical solution for
unambiguous discrimination of two states by using
Lewnstein-Sanpera decomposition is presented. For three linearly
independent states, as the LSD method  leads to a set of coupled
equations which are in general difficult to solve, we embark on
KKT (see Appendix I) method which  makes the problem strongly
easy. Since this condition is necessary and sufficient, then the
answer  will be exactly optimal for unambiguous discrimination.
 \subsubsection{Optimal unambiguous discrimination of two states }
 Suppose that, two pure states $|\psi_{1}\rangle$ and $|\psi_{2}\rangle$
  with arbitrary prior probabilities $\eta_{1}$ and $\eta_{2}$ are given.
 In order to obtain optimal POVM set for these two states by using
 LSD, we use the Lemma $2$ of LSD for two states $|\tilde{\psi}_{1}\rangle$ and
 $|\tilde{\psi}_{2}\rangle$. Let corresponding density matrix in Hilbert
 space $\mathcal{H}$ is identity operator, i.e., $\rho=I$,
and
$\Pi^{\prime}_{1}=\eta_{1}|\tilde{\psi}_{1}\rangle\langle\tilde{\psi}_{1}|$
and
$\Pi^{\prime}_{2}=\eta_{2}|\tilde{\psi}_{2}\rangle\langle\tilde{\psi}_{2}|$.
Then, a pair $(p_{1},p_{2})$ is maximal with respect to $\rho$ and
the pair of operators $\Pi^{\prime}_{1}$ and $\Pi^{\prime}_{2}$
iff
$\rho-p_{1}\eta_{1}|\tilde{\psi}_{1}\rangle\langle\tilde{\psi}_{1}|
-p_{2}\eta_{2}|\tilde{\psi}_{2}\rangle\langle\tilde{\psi}_{2}|\geq
0$. Therefore, from Lemma $2$ of LSD, the maximal pair
$(p_{1},p_{2})$ is given by
\begin{equation} \label{dota}
\begin{array}{c}
  p_{1}=\frac{\tilde{a}_{22}-\sqrt{\frac{\eta_{2}}{\eta_{1}}}|\tilde{a}_{12}|}
{\tilde{a}_{11}\tilde{a}_{22}-|\tilde{a}_{12}|^{2}} \\
  p_{2}=\frac{\tilde{a}_{11}-\sqrt{\frac{\eta_{1}}{\eta_{2}}}|\tilde{a}_{12}|}
{\tilde{a}_{11}\tilde{a}_{22}-|\tilde{a}_{12}|^{2}}, \\
\end{array}
\end{equation}

where
$\tilde{a}_{ij}=\langle\tilde{\psi}_{i}|\tilde{\psi_{j}}\rangle$.
If the condition
\begin{equation}
\frac{|\tilde{a}_{12}|^{2}}{|\tilde{a}_{11}|^{2}+|\tilde{a}_{12}|^{2}}\leq
\eta_{1}\leq
\frac{|\tilde{a}_{11}|^{2}}{|\tilde{a}_{11}|^{2}+|\tilde{a}_{12}|^{2}},
\end{equation}
is hold, then equation (\ref{dota}) is optimal solution.

If $\eta_{1}\leq\eta_{2}$, then the optimal solution reads as
\begin{equation}
p_{1}=0,\quad p_{2}=\frac{1}{\tilde{a}_{22}}.
\end{equation}
If $\eta_{2}\leq\eta_{1}$; then the optimal solution is given by
\begin{equation}
p_{2}=0,\quad p_{1}=\frac{1}{\tilde{a}_{11}}.
\end{equation}
{\bf Examples 4:} For an example we consider following case: Alice
gives Bob a qubit repaired in one of two states \be
|\psi_{1}\rangle=\mid 0\rangle\;\;\;,\;\;\;
|\psi_{2}\rangle=\frac{1}{\sqrt{2}}(\mid 0\rangle+\mid 1\rangle).
\ee Since the states $|\psi_{1}\rangle$ and $|\psi_{2}\rangle$ are
non-orthogonal, there is no measurement that can distinguish them.
In order to obtain optimal POVM, we define dual basis
$|\tilde{\psi}_{j}\rangle$ as follows \be|
\tilde{\psi}_{1}\rangle=(\mid 0\rangle-\mid 1\rangle)\;\;\;,\;\;\;
| \tilde{\psi}_{2}\rangle=\sqrt{2}\mid 1\rangle. \ee Therefore,
pairs $p_{1}$ and $p_2$ are given by
$$p_{1}=p_{2}=\frac{2-\sqrt{2}}{2},$$ and finally the elements of optimal POVM are obtained as
$$
\Pi_{1}=\frac{2-\sqrt{2}}{2}(|0\rangle-|1\rangle)(\langle0|-\langle1|),
\quad \Pi_{2}=(2-\sqrt{2})|1\rangle\langle1|
$$
$$
\Pi_{0}=1-\Pi_{1}-\Pi_{2}.$$
 \subsubsection{Optimal unambiguous discrimination of three linearly independent states }
 Now, we consider three linearly independent normalized
   state vectors $\psi_{1}$ , $\psi_{2}$ and $\psi_{3}$
  with arbitrary prior probabilities $\eta_{1}$, $\eta_{2}$ and $\eta_{3}$ in the three-dimensional complex
 vector space. The KKT conditions for unambiguous
  discrimination of three states are given by
\be\hspace{-1cm}\left\{\begin{array}{c}
     \hspace{-10cm}I-p_{1}|\tilde{\psi}_{1}\rangle\langle\tilde{\psi}_{1}|-p_{2}|\tilde{\psi}_{2}
 \rangle\langle\tilde{\psi}_{2}|-p_{3}|\tilde{\psi}_{3}\rangle\langle\tilde{\psi}_{3}|\geq 0\\
\hspace{-14cm}p_{1}\geq 0, p_{2}\geq 0,p_{3}\geq 0\\
\hspace{-1cm}
(I-p_{1}|\tilde{\psi}_{1}\rangle\langle\tilde{\psi}_{1}|-p_{2}|\tilde{\psi}_{2}
 \rangle\langle\tilde{\psi}_{2}|-p_{3}|\tilde{\psi}_{3}\rangle\langle\tilde{\psi}_{3}|)X=
 X(I-p_{1}|\tilde{\psi}_{1}\rangle\langle\tilde{\psi}_{1}|-p_{2}|\tilde{\psi}_{2}
 \rangle\langle\tilde{\psi}_{2}|-p_{3}|\tilde{\psi}_{3}\rangle\langle\tilde{\psi}_{3}|)=0\\
\hspace{-8cm} z_{1}p_{1}=0,\quad z_{2}p_{2}=0,\quad
z_{3}p_{3}=0,\quad z_{1}\geq
0,z_{2}\geq 0,z_{3}\geq 0\\
Tr(X|\tilde{\psi}_{1}\rangle\langle\tilde{\psi}_{1}|)=z_{1}+\eta_{1},\quad
Tr(X|\tilde{\psi}_{2}\rangle\langle\tilde{\psi}_{2}|)=z_{2}+\eta_{2},Tr(X|\tilde{\psi}_{3}\rangle\langle\tilde{\psi}_{3}|)=
z_{3}+\eta_{3}\quad \eta_{1}\geq 0,\eta_{2}\geq 0,\eta_{3}\geq 0.
 \end{array}\right. \ee
 After some calculation, one can obtain
$$
[(1-p_{2}\tilde{a}_{22})(1-p_{3}\tilde{a}_{33})-p_{2}p_{3}|\tilde{a}_{23}|^{2}][(1-p_{1}\tilde{a}_{11})-
p_{2}\tilde{a}_{12}\sqrt{\frac{\eta_{2}}{\eta_{1}}}-
p_{3}\tilde{a}_{13}\sqrt{\frac{\eta_{3}}{\eta_{1}}}]=0
$$
$$
[(1-p_{1}\tilde{a}_{11})(1-p_{3}\tilde{a}_{33})-p_{1}p_{3}|\tilde{a}_{13}|^{2}][(1-p_{2}\tilde{a}_{22})-
p_{1}\tilde{a}_{12}\sqrt{\frac{\eta_{1}}{\eta_{2}}}-
p_{3}\tilde{a}_{23}\sqrt{\frac{\eta_{3}}{\eta_{2}}}]=0
$$
\begin{equation}
[(1-p_{1}\tilde{a}_{11})(1-p_{2}\tilde{a}_{22})-p_{1}p_{2}|\tilde{a}_{12}|^{2}][(1-p_{3}\tilde{a}_{33})-
p_{2}\tilde{a}_{23}\sqrt{\frac{\eta_{2}}{\eta_{3}}}-
p_{1}\tilde{a}_{13}\sqrt{\frac{\eta_{1}}{\eta_{3}}}]=0.
\end{equation}
With the following conditions
 \be\label{gram4}\left\{
\begin{array}{c}
 \;\; (1-p_{1}\tilde{a}_{11})\geq \frac{p_{2}|\tilde{a}_{12}|^{2}+p_{3}|\tilde{a}_{13}|^{2}}{\tilde{a}_{11}} \hspace{6cm}\\
\;\;(1-p_{2}\tilde{a}_{22})\geq
\frac{p_{1}|\tilde{a}_{12}|^{2}+p_{3}|\tilde{a}_{23}|^{2}}{\tilde{a}_{22}}
\quad, \quad \quad   p_{1}\geq 0,\quad p_{2}\geq 0,\quad p_{3}\geq 0 \\
\;\;(1-p_{3}\tilde{a}_{33})\geq \frac{p_{1}|\tilde{a}_{13}|^{2}+p_{2}|\tilde{a}_{23}|^{2}}{\tilde{a}_{33}},\hspace{6cm} \\
   \end{array}\right. \ee
  the optimal answer is attained as
   $$
p_{1}=\frac{(\tilde{a}_{22}\tilde{a}_{33}-|\tilde{a}_{23}|^{2})-\sqrt{\frac{\eta_{2}}
{\eta_{1}}}(\tilde{a}_{21}\tilde{a}_{33}-\tilde{a}_{31}\tilde{a}_{23})+\sqrt{\frac{\eta_{3}}{\eta_{1}}}
(\tilde{a}_{21}\tilde{a}_{32}-\tilde{a}_{22}\tilde{a}_{31})}
{\tilde{a}_{11}(\tilde{a}_{22}\tilde{a}_{33}-|\tilde{a}_{23}|^{2})
-a_{12}(\tilde{a}_{21}\tilde{a}_{33}-\tilde{a}_{31}\tilde{a}_{23})+\tilde{a}_{13}
(\tilde{a}_{21}\tilde{a}_{32}-\tilde{a}_{22}\tilde{a}_{31})}
$$
$$
p_{2}=\frac{(\tilde{a}_{11}\tilde{a}_{33}-|\tilde{a}_{13}|^{2})-\sqrt{\frac{\eta_{1}}
{\eta_{2}}}(\tilde{a}_{12}\tilde{a}_{33}-\tilde{a}_{32}\tilde{a}_{13})+
\sqrt{\frac{\eta_{3}}{\eta_{2}}}(\tilde{a}_{11}\tilde{a}_{32}-\tilde{a}_{31}\tilde{a}_{12})}
{\tilde{a}_{11}(\tilde{a}_{22}\tilde{a}_{33}-|\tilde{a}_{23}|^{2})
-\tilde{a}_{12}(\tilde{a}_{21}\tilde{a}_{33}-\tilde{a}_{31}\tilde{a}_{23})+\tilde{a}_{13}
(\tilde{a}_{21}\tilde{a}_{32}-\tilde{a}_{22}\tilde{a}_{31})}
$$
\begin{equation}
p_{3}=\frac{(\tilde{a}_{11}\tilde{a}_{22}-|\tilde{a}_{12}|^{2})+\sqrt{\frac{\eta_{1}}
{\eta_{3}}}(\tilde{a}_{12}\tilde{a}_{23}-\tilde{a}_{22}\tilde{a}_{13})-
\sqrt{\frac{\eta_{2}}{\eta_{3}}}(\tilde{a}_{11}\tilde{a}_{23}-\tilde{a}_{21}\tilde{a}_{13})}
{\tilde{a}_{11}(\tilde{a}_{22}\tilde{a}_{33}-|\tilde{a}_{23}|^{2})
-\tilde{a}_{12}(\tilde{a}_{21}\tilde{a}_{33}-\tilde{a}_{31}\tilde{a}_{23})+
\tilde{a}_{13}(\tilde{a}_{21}\tilde{a}_{32}-\tilde{a}_{22}\tilde{a}_{31})}.
\end{equation}
    If the conditions (4-82) do not
   satisfied, then we search another optimal solution. If
\begin{equation}
\frac{|\tilde{a}_{23}|^{2}}{|\tilde{a}_{22}|^{2}+|\tilde{a}_{23}|^{2}}\leq
\eta_{2}\leq
\frac{|\tilde{a}_{22}|^{2}}{|\tilde{a}_{22}|^{2}+|\tilde{a}_{23}|^{2}},
\end{equation}
is hold, the optimal solation  is given by \be\label{gram4}\left\{
\begin{array}{c}
\;\;\hspace{-2cm}p_{1} =0\\
 \;\;p_{2}=\frac{|\tilde{a}_{33}|-\sqrt{\frac{\eta_{3}}{\eta_{2}}}|\tilde{a}_{23}|}
 {|\tilde{a}_{22}||\tilde{a}_{33}|-|\tilde{a}_{23}|^{2}} \\
\;\;p_{3}=\frac{|\tilde{a}_{22}|-\sqrt{\frac{\eta_{2}}{\eta_{3}}}|\tilde{a}_{23}|}
 {|\tilde{a}_{22}||\tilde{a}_{33}|-|\tilde{a}_{23}|^{2}}. \\

   \end{array}\right.\ee
   If
   \begin{equation}
\frac{|\tilde{a}_{13}|^{2}}{|\tilde{a}_{11}|^{2}+|\tilde{a}_{13}|^{2}}\leq
\eta_{1}\leq
\frac{|\tilde{a}_{11}|^{2}}{|\tilde{a}_{11}|^{2}+|\tilde{a}_{13}|^{2}},
\end{equation}
is hold, the optimal answer is given by \be\label{gram4}\left\{
\begin{array}{c}
 \;\;p_{1}=\frac{|\tilde{a}_{33}|-\sqrt{\frac{\eta_{3}}{\eta_{1}}}|\tilde{a}_{13}|}
 {|\tilde{a}_{11}||\tilde{a}_{33}|-|\tilde{a}_{13}|^{2}} \\
 \;\;\hspace{-2cm}p_{2} =0\\
\;\;p_{3}=\frac{|\tilde{a}_{11}|-\sqrt{\frac{\eta_{1}}{\eta_{3}}}|\tilde{a}_{13}|}
 {|\tilde{a}_{11}||\tilde{a}_{33}|-|\tilde{a}_{13}|^{2}}. \\

   \end{array}\right.\ee
   If the condition
\begin{equation}
\frac{|\tilde{a}_{12}|^{2}}{|\tilde{a}_{11}|^{2}+|\tilde{a}_{12}|^{2}}\leq
\eta_{1}\leq
\frac{|\tilde{a}_{11}|^{2}}{|\tilde{a}_{11}|^{2}+|\tilde{a}_{12}|^{2}},
\end{equation}
is satisfied, the optimal solution reads as
\be\label{gram4}\left\{
\begin{array}{c}
 \;\;p_{1}=\frac{\tilde{a}_{22}-\sqrt{\frac{\eta_{2}}{\eta_{1}}}|\tilde{a}_{12}|}
{\tilde{a}_{11}\tilde{a}_{22}-|\tilde{a}_{12}|^{2}} \\
 \;\;p_{2} =\frac{\tilde{a}_{11}-\sqrt{\frac{\eta_{1}}{\eta_{2}}}|\tilde{a}_{12}|}
{\tilde{a}_{11}\tilde{a}_{22}-|\tilde{a}_{12}|^{2}}.\\
\;\;\hspace{-2cm}p_{3}=0 \\
   \end{array}\right.\ee
   If none of the above conditions is hold, in this case two of $p_{i}$
   vanish and the optimal POVM will be the same as Von Numan measurement.

{\bf Example 5:}
  In this example we consider reciprocal
 independent states $\{\tilde{\psi}_{i}\}_{i=1}^{N}$ such that
$$\langle\tilde{\psi}_{i}\mid\tilde{\psi}_{j}\rangle=a\quad {\mathrm{for}} \quad i\neq j$$
\begin{equation}
\langle\tilde{\psi}_{i}\mid\tilde{\psi}_{j}\rangle=1\quad
{\mathrm{for}} \quad i=j,
\end{equation}

 Thus, for optimal distinguishing of independent vectors that are
prepared with equal probabilities, we minimize the inconclusive
result $\Pi_{0}$ given by \be
\Pi_{0}=I-\sum_{i=1}^{N}p_{i}\mid\tilde{\psi}_{i}\rangle\langle\tilde{\psi}_{i}\mid.
\ee
 Now using LS's theorem and some analytical calculations we
show that all of $p_i$'s are equal, i.e., all of the elements of
POVM possess equal probabilities, i.e., \be
p_{1}=p_{2}=...=p_{N}=\frac{1}{1+a(N-1)}. \ee Similar to GU
states, one can prove that the optimal distinguishability
corresponds to equal measurement probabilities $p_i=p$ for all $i$
such that $p$ is equal to the inverse of maximum eigenvalue of the
corresponding frame operator. In order to prove this fact, we
define frame operator as \be
S=\sum_{i=1}^{N}\mid\tilde{\psi}_{i}\rangle\langle\tilde{\psi}_{i}\mid,
\ee such that \be S_{kl}=\sum_{i=1}^{N}\langle
k\mid\tilde{\psi}_{i}\rangle\langle\tilde{\psi}_{i}\mid l\rangle.
\ee Then, $S=AA^{\dagger}$ with $A_{ki}=\langle
k\mid\tilde{\psi}_{i}\rangle$. On the other hand, the Gramm matrix
is defined as follows \be \tilde{G}=A^{\dagger}A. \ee Again, one
can easily show that $S$ and $G$ have equal eigenvalues, thus we
evaluate eigenvalues of Gramm matrix . The Gramm matrix
$\tilde{G}$ can be written as \be \tilde{G}=I+a(C-1), \ee such
that its eigenvalues are given by\be \left\{\begin{array}{c}
1+a(N-1)\\ 1-a,
\end{array}\right. \ee where $a<1$. Therefore, maximum eigenvalue of
frame operator is equal  to $1+a(N-1)$ and thus $p$ is given by
$(1+a(N-1))^{-1}$.
\subsection{Discrimination of quantum states using approximated linear programming}
As solving the problem analytically is so hard, approximated
methods are useful for unambiguous discrimination of $N$ linearly
independent quantum states. The simplex method is the easiest way
of solving it. The simplex algorithm is a common algorithm used to
solve an optimization problem with a polytope feasible region,
such as a linear programming problem. It is an improvement over
the algorithm to test all feasible solution of the convex feasible
region and then choose the optimal feasible solution. It does this
by moving from one vertex to an adjacent vertex, such that the
objective function is improved. This algorithm still guarantees
that the optimal point will be discovered. In addition, only in
the worst case scenario all vertices will be tested. Here,
considering the scope of this paper, a complete treatment of the
simplex algorithm is unnecessary; for a more complete treatment
please refer to any LP text such as \cite{boyd,chong}. In the
following, we give some examples and compare the result with
analytical solution. Suppose that, the quantum system is prepared
in one of the two pure states $|\psi_{1}\rangle$ and
$|\psi_{2}\rangle$
  with arbitrary prior probabilities  $\eta_{1}$ and $\eta_{2}$.
 To unambiguously discrimination, one should calculate the optimal $p_i$'s by solving equation
 $(\ref{fise1})$. According the equation ($\ref{fise1}$), the three extremal
points \be\begin{array}{ccc}
\left\{\begin{array}{c}p_{1}=0\\p_{2}=1-|\langle\psi_{1}|\psi_{2}\rangle|^{2}\end{array}\right.,
&
\left\{\begin{array}{c}p_{2}=0\\p_{1}=1-|\langle\psi_{1}|\psi_{2}\rangle|^{2}\end{array}\right.,
&
\left\{\begin{array}{c}p_{1}=p_{2}=1-|\langle\psi_{1}|\psi_{2}\rangle|.\end{array}\right.\end{array}
\ee together with the origin ($p_{1}=p_{2}=0$) form a polygon
which surrounds the feasible region. Since the feasible region is
not linear(see Figure $1$) while the polygon is linear, so we use
approximated simplex method to find optimal answer. For three
linearly independent normalized state vectors $|\psi_{1}\rangle$ ,
$|\psi_{2}\rangle$ and $|\psi_{3}\rangle$ with arbitrary prior
probabilities $\eta_{1}$, $\eta_{2}$ and $\eta_{3}$ in the
three-dimensional complex vector space, there is seven extremal
points. These points together with the origin form a polygon. The
optimal solutions, resulting from the LP method, are consistent to
that ones obtained analytically in sections 3 and 4 with high
accuracy.
\par
Since obtaining analytical solution for more than three states is
difficult, the approximated LP method seems to be  useful for
optimal unambiguous discrimination of $N$ linearly independent
states. This method, not only enhance the speed of calculation but
also is done with high precision.
\section{Conclusion}
Here in this work by reducing the theory of the semi-definite
programming to a linear programming one with a feasible region of
polygon type which can be solved via simplex method, we have been
able to obtain optimal measurements to unambiguous discrimination of
an arbitrary number of pure linearly independent quantum states and
using the close connection between the Lewenstein-Sanpera
decomposition and semi-definite programming, we have been able to
obtain the optimal positive operator valued measure  for some of the
well known  examples via Lewenstein-Sanpera decomposition method.
Unambiguous discrimination of mixed states is under investigation.

\vspace{1cm} \setcounter{section}{0}
 \setcounter{equation}{0}
 \renewcommand{\theequation}{I-\arabic{equation}}
{\Large{Appendix I:}}\\
\section{Karush-Kuhn-Tucker (KKT) theorem: }
 Assuming that functions $g_{i}$, $h_{i}$
are differentiable and that strong duality holds, there exists
vectors $\zeta \in R^{k}$, and $y \in R^{m}$, such that the
 gradient of dual Lagrangian $L(x^{\ast}, \zeta^{\ast}, y^{\ast}) = f(x^{\ast}) +
\sum_{i} \zeta_{i}^{\ast}h_{i}(x^{\ast}) + \sum_{i}
y_{i}^{\ast}g_{i}(x^{\ast})$ over $x$ vanishes at $x^{\ast}$ :

$$h_{i}(x^{\ast}) = 0\quad(\mathrm{primal\ feasible})
$$
$$
 g_{i}(x^{\ast}) \leq 0\quad(\mathrm{primal\ feasible})
 $$
 $$
  y^{\ast}_{i} \geq 0\quad(\mathrm{dual \ feasible})
  $$
  $$
  y^{\ast}_{i}
g_{i}(x^{\ast}) = 0
$$
 \begin{equation}
\bigtriangledown f(x^{\ast}) +\sum_{i}\zeta^{\ast}_{i}
\bigtriangledown h_{i}(x^{\ast}) +\sum_{i}y^{\ast}_{i}
\bigtriangledown g_{i}(x^{\ast}) = 0.
  \end{equation}
Then $x^{\ast}$ and $(\zeta^{\ast}_{i}, y^{\ast}_{i})$ are primal
and dual optimal with zero duality gap. In summary, for any convex
optimization problem with differentiable objective and constraint
functions, the points which satisfy the KKT conditions are primal
and dual optimal, and have zero duality gap. Necessary KKT
conditions satisfied by any primal and dual optimal pair and for
convex problems, KKT conditions are also suffcient. If a convex
optimization problem with differentiable objective and constraint
functions satisfies Slater's condition, then the KKT conditions
provide necessary and suffcient conditions for optimality:
Slater's condition implies that the optimal duality gap is zero
and the dual optimum is attained, so x is optimal if and only if
there are $(\zeta_{i}^{\ast}, y^{\ast}_{i})$ such that, together
with $x$ satisfy the KKT conditions.
\subsection{Slater's condition: }

Suppose $x^{\ast}$ solves
 \begin{equation}
  {\mathrm{minimize}} \quad f(x)  g_{i}(x) \geq b_i , i = 1,
...,m,
 \end{equation}
and the feasible set is non empty. Then there is a non-negative
vector $\zeta$ such that for all $x$

 \begin{equation}
 L(x, \zeta) = f(x) + \zeta^{T}(b - g(x)) \leq f(x^{\ast}) =
L(x^{\ast}, \zeta).
 \end{equation}
  In addition, if $f(.), g_i(.) , i = 1, ...,m$
are continuously differentiable, then
 \begin{equation}
\frac{\partial f(x^{\ast})}{\partial (x_{j})}-\zeta\frac{\partial
g(x^{\ast})}{\partial (x)}=0.
 \end{equation}
In the spatial case the vector $x$ is a solution of the linear
program
$$
{\mathrm{minimize}} \quad c^{T}x
$$
\begin{equation}
{\mathrm{s.t}} \quad Ax = b  x \geq 0,
\end{equation}
if and only if there exist vectors $\zeta\in R^{k}$, and $y \in
R^{m}$ for which the following conditions hold for $(x, \zeta, y)
= (x^{\ast}, \zeta^{\ast}, y^{\ast})$
\begin{equation}
 A^{T}\zeta + y = c \quad Ax = b \quad x_{i} \geq
0; y_{i} \geq 0; \; x_{i}y_{i} = 0 , i = 1, ...,m.
\end{equation}
A solution $(x^{\ast}, \zeta^{\ast}, y^{\ast})$ is called strictly
complementary, if $x^{\ast} + y^{\ast} > 0$, i. e., if there
exists no index $i\in {1, ...,m}$ such that $x^{\ast}_{i} =
y^{\ast}_{i} = 0$.

\newpage
{\bf Figure Captions}

 {\bf Figure-1:} Feasible region for unambiguous discrimination of two linearly independent
 states with $a_{12}=<\psi_{1}\mid\psi_{2}>$ is showed by shadow region and approximately feasible region is showed by a polygon.

 { \bf Figure-2:} Unambiguous discrimination of two pure states in Bloch sphere.

{\bf Figure-3:} Feasible region for unambiguous discrimination of
three linearly independent
 states with $a_{12}=a_{13}=\frac{1}{3} \quad and\quad
 a_{23}=\frac{1}{9}$.

{\bf Figure-4:} Feasible region for unambiguous discrimination of
three linearly independent with $a_{12}=a_{13}=\frac{2}{\sqrt {3}}
\quad and\quad a_{23}=\frac{1}{2}$
 states.

\end{document}